# Formation of Planets by Hydrogravitational Dynamics

Carl H. Gibson [1,2]

[1] University of California San Diego, La Jolla, CA 92093-0411, USA
[2] cgibson@ucsd.edu, http://sdcc3.ucsd.edu/~ir118

and

Rudolph E. Schild [3,4]

[3] Center for Astrophysics, 60 Garden Street, Cambridge, MA 02138, USA
[4] rschild@cfa.harvard.edu

## ABSTRACT

From hydro-gravitational cosmology, hydrogen-helium gas planets fragmented at the plasma to gas transition 300,000 years after the big bang in million-star-mass clumps. Stars may form in the clumps by mergers of the planets to make globular star clusters. Star-less clumps persist as the dark matter of galaxies as observed by Schild in 1996 using quasar microlensing, and as predicted by Gibson in 1996 using fluid mechanics. Massive plasma structures, at $10^{46}$ kg proto-galaxy-cluster-mass, fragment at 30,000 years when photon-viscous forces match gravitational fragmentation forces at the horizon scale $ct$ of the expanding universe, where $c$ is the speed of light and $t$ is the time. Spinning proto-super-cluster-void and proto-galaxy-void boundaries expand at sound speeds $c/3^{1/2}$ producing weak turbulence and linear-clusters of gas-proto-galaxies that are fossils of turbulent-plasma vortex lines. Hubble-space-telescope images of the most distant galaxies support this Gamov 1951 prediction. Vortex spin axes inferred from microwave background anisotropies are interpreted as evidence of a turbulent big bang. A cosmic distribution of life is attributed to hot water oceans of the interacting hydrogen planets seeded by the first chemicals.

## 1. Introduction

The standard (concordance) cold dark matter model of cosmology is based on the Jeans 1902 theory of gravitational instability of a gas nebula, where numerous unwarranted assumptions are made to simplify the conservation laws of fluid mechanics  (Jeans 1902).  Jeans assumed inviscid Euler momentum equations with linear perturbation stability analysis, thus neglecting crucially important viscous forces and turbulence forces.  He also neglected any effects of diffusivity on gravitational structure formation, unaware of a massive population of super-diffusive neutrinos.   The resulting acoustics equations were then solved by neglecting density (the Jeans swindle) giving the Jeans length scale $L_J = V_S/(\rho G)^{1/2}$ as the single criterion for gravitational instability, where $V_S$ is the speed of sound, $\rho$ is density and G is Newton's gravitational constant.

Density wavelengths smaller than the Jeans scale were incorrectly assumed by Jeans 1902 to be stabilized by pressure forces.  This is also a fatal error of the concordance model.  Those density wavelengths larger than $L_J$ are incorrectly taken to be unstable linear modes that may grow in amplitude without limit.  Large errors result from the Jeans 1902 criterion, which fails to predict the formation of primordial-gas planets at the plasma-gas transition (Gibson 1996), and fails to predict fragmentation of supercluster, cluster, and galaxy mass objects in the plasma. Significant errors result from a lack of understanding of turbulence and fossil turbulence, since the big bang is a result of a turbulence instability (Gibson 2005) and the boundary conditions of the evolving universe are set by the fossils of big bang turbulence (Gibson 2004).  Numerous manifestations of big bang fossil vorticity turbulence have been documented (Schild and Gibson  2008, Gibson 2010ab).  The existence of an "axis of evil" preferred direction on the sky clearly shows a first order departure from the standard cosmological model.

Because the Jeans scale is larger than the horizon scale $L_H = ct$ during the plasma epoch, no plasma structures should form during this period 3,000 to 300,000 years ($10^{11}$ to $10^{13}$ seconds) after the big bang event by the Jeans 1902 acoustic criterion. The baryonic material (H and He plasma) is necessarily a small fraction, $\sim 4\%$, of the total mass required for a "flat" (neither open nor closed) universe (Peebles 1993).

According to "cold-dark-matter" CDM theories, the non-baryonic dark matter component with its unknown weakly-collisional particles is arbitrarily assumed to be cold to reduce the CDM Jeans scale $L_{JCDM}$ below the horizon $L_H$ during the plasma epoch, permitting condensation as "seeds", or cold-dark-matter CDM halos. Observational tests of the CDM model fail badly in the local group of galaxies (Kroupa et al. 2010).

Such CDM halos were imagined to "hierarchically cluster" HC to much larger masses over time, producing massive gravitational potential wells into which the baryonic-dark-matter BDM could fall and form the first stars in a few hundred million years of dark ages. The resulting CDMHC model is standard in present day astronomy, astrophysics and cosmology.

Is the standard CDMHC model true? Can viscosity be neglected in the plasma epoch? Can turbulence[1] be neglected? Can fossil turbulence[2] be neglected? If the non-baryonic-dark-matter NBDM is nearly collisionless, can its diffusivity be neglected? No, clearly not. Whatever it is, NBDM is super-diffusive and cannot condense, as assumed by CDMHC, to guide the structure formation of the BDM. The reverse is true (Gibson 1996). Even though its mass is smaller, the sticky, collisional, baryonic matter forms galaxy and galaxy cluster structures that gravitationally guide the non-baryonic material to form large cluster and supercluster halos. When fluid mechanical constraints are imposed an entirely different cosmology emerges, termed hydrogravitational dynamics HGD.

As a natural part of the discussion of the two theories of cosmology, the question of a permanent form of anti-gravity arises. Anti-gravity is needed in any big bang theory beginning at Planck densities. This is why Einstein was forced to include a cosmological constant Λ in his general relativity equations (Peacock 2000, p15). Anti-gravitational forces occur naturally in the turbulent HGD big bang model due to turbulence stresses and gluon viscosity, and vanish at the end of the inflation epoch.

---

[1] Turbulence is defined as an eddy-like state of fluid motion where the inertial-vortex forces of the eddies are larger than any other forces that tend to damp the eddies out. By this definition, turbulence always cascades from small scales to large.

[2] Fossil turbulence is defined as a perturbation in any hydrophysical field produced by turbulence that persists after the fluid is no longer turbulent at the scale of the perturbation.

Thus a permanent "dark energy" cosmological constant Λ is unnecessary. A much more plausible candidate for the NBDM is provided by neutrinos (Nieuwenhuizen 2009). Nieuwenhuizen estimates a neutrino mass sufficient to exceed the baryonic matter by a factor of six, and possibly the factor of thirty needed to flatten the universe. Because turbulence creates entropy, HGD cosmology predicts a closed universe with non-baryonic dark matter mass in this 6-30 times baryonic range.

In the following, we first summarize the hydrogravitational dynamics theory that leads to the formation of primordial planets as the dark matter of galaxies. We contrast HGD predictions with those of standard CDMHC cosmologies, and provide discussion of results and conclusions.

## 2. Comparison of HGD cosmology with ΛCDMHC cosmology

Figure 1abcde summarizes HGD cosmology. Turbulence dominates the big bang event shown at the right of Fig. 1a until the kinematic viscosity increases from Planck values of order $\nu_P = L_P \times c \sim 10^{-27}$ m$^2$ s$^{-2}$, to gluon-viscous values that are much larger, where $L_P$ is the Planck length scale and c is the speed of light. Reynolds numbers of the turbulent combustion process monotonically increase drawing energy from the vacuum by negative turbulence stresses (Gibson 2010).

Much larger negative viscous stress occur when the spinning turbulent fireball cools from $10^{32}$ K to $10^{28}$ K, so the quark-gluon plasma can form. Gluons transfer momentum over larger collision distances than the Planck scale, increasing the viscosity and negative stresses proportionately. An exponential inflationary event is triggered, with displacement velocities up to $\sim 10^{25}$ $c$, increasing the mass-energy of the universe to values $\sim 10^{97}$ kg, compared to $\sim 10^{53}$ kg visible in our present horizon $L_H = ct$. Because the turbulent temperature fluctuations of the big bang are stretched far beyond the horizon scales $L_H$, they become the first fossil temperature turbulence (Gibson 2004). Vortex lines produced at Planck scales persist throughout the inflationary expansion to become the first fossil vorticity turbulence.

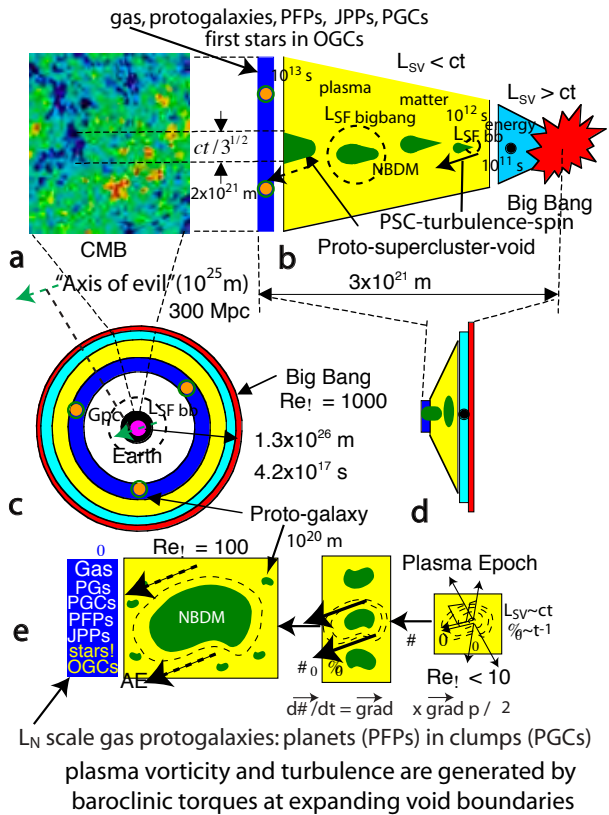

Fig. 1. Hydrogravitational dynamics cosmology (Gibson, 2010). In **a.** CMB temperature anomaly patterns reflect protosupercluster (PSC) and protosuperclustervoid patterns at the sonic scale $ct/3^{1/2}$ of the voids formed at $10^{12}$ seconds when the viscous-gravitational scale $L_{SV}$ first matched the increasing horizon scale $L_H$. Vorticity **e.** produced at void boundaries determines the morphology of plasma protogalaxies, which fragment into proto-globulular-star-clusters PGC clumps of primordial-fog-particle PFP planets at transition to gas.

A mass of $\sim 10^{16}$ kg is produced by the big bang event in $\sim 10^{-27}$ sec, giving a power of $10^{60}$ watts. A mass of $\sim 10^{97}$ kg is produce by inflation in $\sim 10^{-33}$ sec, giving a power of $10^{147}$ watts. Nucleosynthesis gives H, $^4$He and electrons in $\sim 10^3$ sec. Mass exceeds energy at $\sim 10^{11}$ sec. Gravitational structure formation cannot occur until the plasma epoch, when the horizon scale $L_H$ exceeds the viscous Schwarz scale $L_{SV}$ at time $\sim 10^{12}$ sec. As shown in Fig. 1e, the first gravitational structures were protosuperclustervoids due to fragmentation at density minima. Mass scales of the fragments $\sim 10^{46}$ kg match supercluster masses. Weak turbulence produced by torques at spinning void boundaries has been detected by Sreenivasan and Bershadskii from the statistics of CMB temperature anisotropies (Gibson 2010).

The standard $\Lambda$CDMHC cosmology attempts to make gravitational structures in the plasma epoch as shown in Figure 2 (top). The non-baryonic dark matter is assumed to be cold to reduce the Jeans scale to less than the scale of causal connection $L_H$. Cold dark matter CDM seeds are assumed to condense and hierarchically cluster HC to from gravitational potential wells to collect the baryonic plasma. Accoustic oscillations of the plasma within such potential wells accounts for the prominent acoustic peak observed in CMB temperature anisotropies.

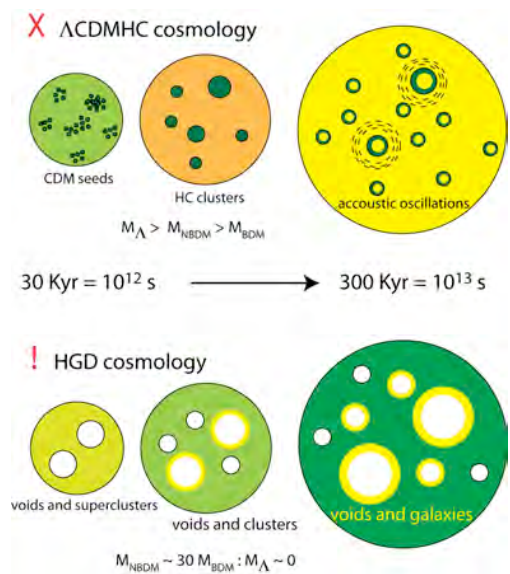

Fig. 2. Contrast between $\Lambda$CDMHC and HGD models of structure formation during the plasma epoch. At the top, cold dark matter CDM seeds condense at the Jeans scale and then hierarchically cluster HC to form potential wells that collect acoustically oscillating plasma. At the bottom, hydrogravitational dynamics HG predicts a sequence of fragmentations from supercluster to galaxy scales, controlled by viscous and turbulence forces (Gibson 2010).

Figure 2 (bottom) shows the HGD structure formation scenario. Even though the NBDM mass greatly exceeds that of the baryonic plasma, it cannot condense or cluster until after the plasma epoch because it is so weakly collisional (Gibson 2000). Fossils of the turbulent rate-of-strain, density, and spin at the $t_0 = 10^{12}$ sec time of first structure persist to this day.

Figure 3 (top) contrasts ΛCDMHC cosmologies with the HGD cosmology Figure 3 (bottom) during the early years of the gas epoch between $10^{13}$ sec and $10^{16}$ sec.

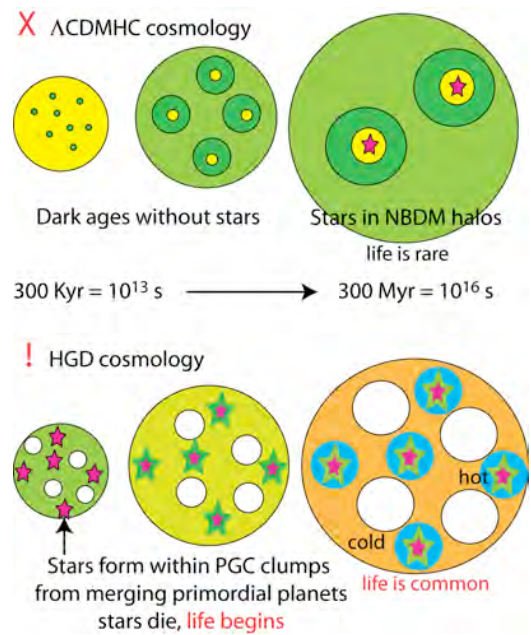

Fig. 3. Contrast between ΛCDMHC and HGD models of structure formation during the early years of the gas epoch beginning at $10^{13}$ sec. The first stars appear after $10^{16}$ sec according to ΛCDMHC (top), ending 300 Myr of dark ages. The first stars and first chemicals occur quite promptly according to HGD cosmology (bottom), within the fossil freefall time $t_0 = 10^{12}$ sec. Jeans mass protoglobularstarcluster PGC clumps of primordial-fog-particle PFP planets form at the transition time $10^{13}$ sec., setting the stage for the biological big bang (Gibson, Schild and Wickramasinghe 2010) beginning at $6 \times 10^{13}$ sec when planetary oceans first condense.

All stars form within PGC clumps of PFP planets by a sequence of mergers of the hydrogen-helium gas planets. Old globular star clusters OGC have the baryonic density $\rho_0 = 4 \times 10^{-17}$ kg m$^{-3}$ of the time of first structure $t_0$. This is also the density of protogalaxies formed with Nomura scale $10^{20}$ meters and the Nomura morphology determined by numerical simulations of weak turbulence, as shown in Figure 4.

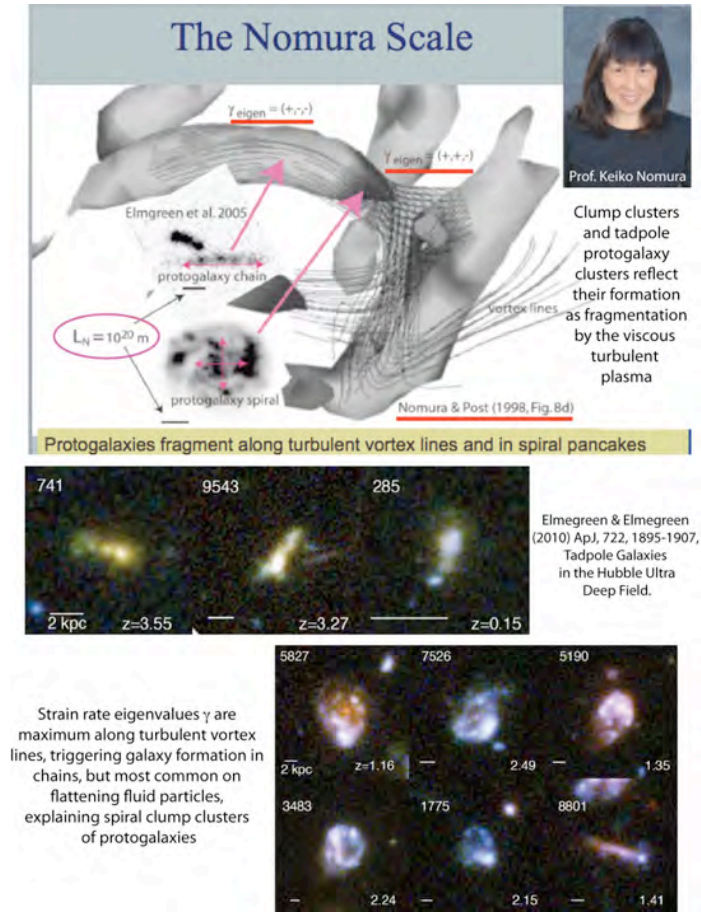

Fig. 4.  Tadpole and clump cluster protogalaxies observed by Elmegreen & Elmegreen (2010) compared to the morphology of weak turbulence simulated by Nomura & Post (1998).

Direct numerical simulations of weak turbulence (Nomura & Post 1998) confirm the mathematical expectation that most spherical turbulent fluid particles flatten into pancakes with rate-of-strain eigenvalues $\gamma_{eigen}$ (+,+,-).  The unanticipated morphology is that these form spiral structures at the base of vortex lines with $\gamma_{eigen}$ (+,-,-), as shown at the top of Fig. 4.  Protogalaxies fragmenting at the end of the plasma epoch maintain this Nomura morphology, as shown by the linear and spiral clusters of galaxies detected in the Hubble Space Telescope Ultra Deep Field images.

Horizontal bars of scale 2 kpc are shown showing the bright clumps are close to the Nomura viscous length scale of the late plasma epoch, at $L_N = 10^{20}$ meters.

This high density $\rho_0 = 4 \times 10^{-17}$ kg m$^{-3}$ of globular clusters and protogalaxies exceeds the average galaxy density by four orders of magnitude. Stars in OGCs are typically small and long lived, supporting the HGD scenario that the transition from plasma to gas was extremely gentle. First stars formed in the standard cosmology scenario of Fig. 3 (top) are claimed to be such massive superstars that when they condense from the turbulent gas clouds collected by the massive assembled CDM halos Fig. 3 (top right) they immediately detonate to re-ionize the universe. The re-ionization hypothesis is unnecessary to explain the lack of hydrogen observed in distant quasar spectra. It is unsupported by observations intended to detect the enormous radiation events of the superstar supernovae. It never happened. According to the HGD cosmology of Fig. 3 (bottom) the hydrogen planets in their PGC clumps have become the dark matter of galaxies by freezing. Figure 5 shows the HGD model of spiral galaxy formation. The non-baryonic dark matter diffuses to larger scales than the baryonic dark matter halos, and provides a small fraction of the density within the PGC baryonic dark matter halo.

As shown in Fig. 5 (top) the plasma protogalaxies promptly fragment at the PGC Jeans mass of a million solar masses, each containing a trillion primordial gas fog particle PFPs with Earth mass. The time of fragmentation is determined by the large density $\rho_0$ of the protogalaxies, so the PGC/PFP formation time $\tau_g$ is significantly smaller than $t$. The same rapid time of formation applies to mergers of PFPs to make larger planets and finally stars. Thus the first star of HPD cosmology should appear in a PGC near the core of a protogalaxy at a time not much different than the plasma to gas transition time $10^{13}$ seconds, as shown in Fig. 5 (top right). Strong viscous forces of the gas planets requires the mergers of planets to be gentle, so the first stars formed will be small and long lived, as observed in old globular star clusters OGCs. Such mergers should be mostly binary, leading to binary planets and binary stars, as observed.  Numerous planet fragments are collected gravitationally as comets and meteors by the larger objects, including biochemical information produced as living organisms form when the first oceans are seeded with the first chemicals (Gibson, Schild and Wickramasinghe 2010).

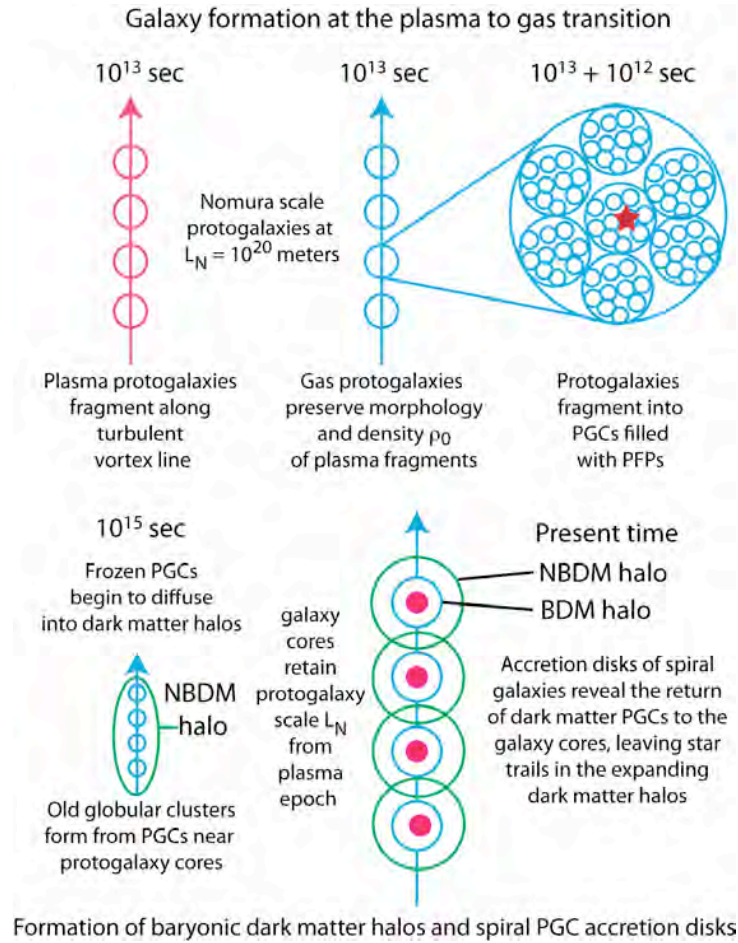

Fig. 5.  Galaxy formation at the plasma to gas transition (top) at time $10^{13}$ sec.  Formation of baryonic dark matter halos by diffusion of freezing baryonic dark matter PGCs out of the $L_N$ scale galaxy cores (bottom).

The model of spiral galaxy formation of Fig. 5 is supported by the compound image of nearby galaxy M81, shown in Figure 6.  The central core of old stars matches the $L_N$ scale of the original protogalaxy from which PGCs diffused to form the $10^{22}$ meter diameter dark matter halo.  Those clumps of PGCs that escaped from the core have small direct collision probability for their frozen planets, which makes them strongly diffusive, but large collision probability for tidal interactions, which tiggers star formation and friction.  Infrared images of protostars forming in the wake of PGC centers of gravity can be seen in release images of the Milky Way disk from the Planck space telescope and the Herschel space telescope (Gibson, Schild & Wickramasinghe 2010, figs. 11 & 12).

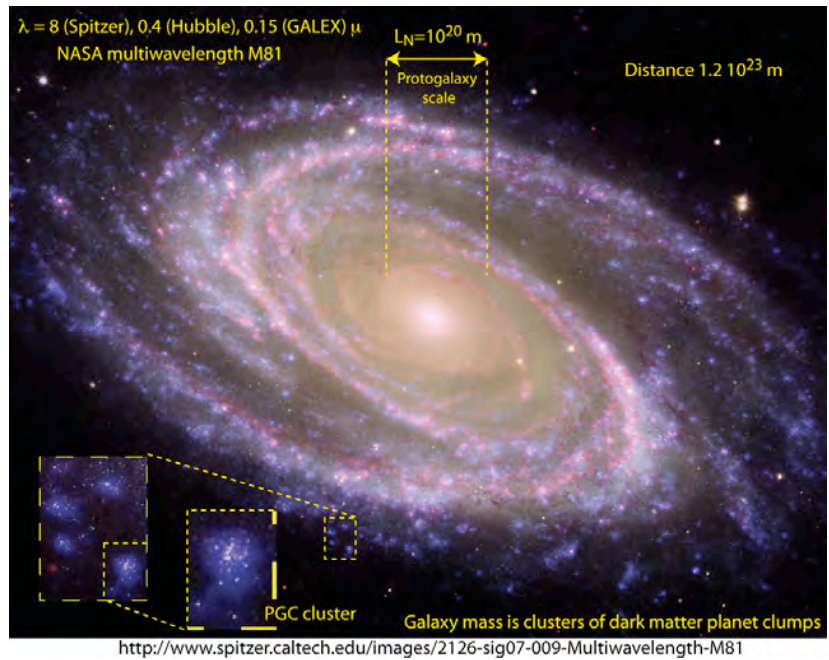

Fig. 6.  A multiwavelength image of nearby galaxy M81 ($1.2 \times 10^{23}$ m) is shown from the infrared telescope Spitzer at 8 microns, the Hubble space telescope at 0.4 microns, and the ultraviolet 0.15 micron telescope GALEX.  Star trails of the merging objects are revealed in the frictional PGC accretion disk of M81.  The baryonic dark matter of the spiral galaxy accretion disk is revealed to be clusters of PGCs (lower left).  Individual PGCs appear as bright dots, probably turbulent O-B star complexes triggered into formation by tidal forces.

Temperatures detected by the infrared telescopes Planck and Herschel match triple point and critical temperatures expected for boiling and freezing hydrogen planets as they merge, collide and re-freeze as expected in the HGD scenario for star formation by accretional cascades from PFP mass to star mass within PGC clumps.

## 3. Discussion

The standard model ΛCDMHC is physically untenable and observationally unsupported.  As shown in Fig. 2 (top) the scenario requires gravitational condensation of a nearly collisionless fluid: cold dark matter CDM.  How can a nearly collisionless fluid condense, and how can its condensates merge to form stable CDM halos?  No matter how cold and motionless the CDM particles are initially, they will not be cold for long, and even if they are merged somehow, they cannot hierarchically cluster HC to form larger mass halos.

Imagine a sphere of perfectly cold dark matter, where its particles are initially completely motionless, as shown in Figure 7.

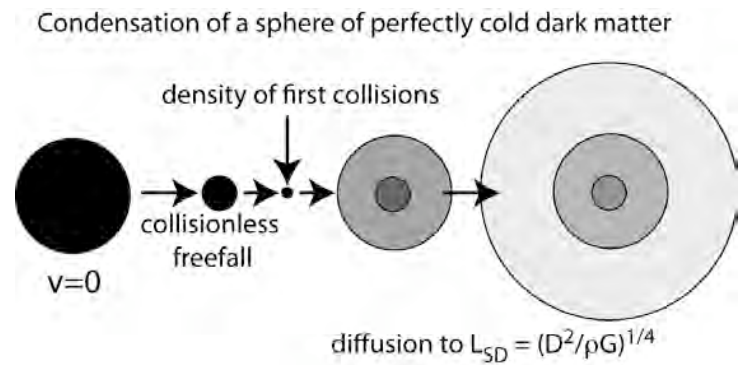

Fig. 7.  Instability of weakly collisional cold dark matter to diffusion.  A sphere of motionless cold dark matter particles would collapse to a density where collisions occur, so the particle motions must randomize and thermalize.  The sphere will grow to the Schwarz diffusive scale, where the diffusion velocity matches the gravitational velocity (Gibson 2000).

As shown in the Fig. 7 counterexample, CDM condensations are physically impossible.  So are mergers of CDM condensates by hierarchical clustering HC.

Gluon viscous forces terminate the big bang, fossilizing turbulent big bang temperature fluctuations by inflation beyond the scale of causal connection $ct$.  Gravitational structure begins in the plasma epoch when the Schwarz viscous scale $L_{ST}$ matches $ct$ at time $10^{12}$ sec.  $L_{ST}$ is $(\gamma\nu/\rho G)^{1/2}$, where $\gamma$ is the rate-of-strain $\sim t^{-1}$, $\nu$ is the kinematic viscosity, $\rho$ is the density and G is Newton's constant. The viscosity during the plasma epoch occurs by photon collisions with free electrons. The collision cross section for such events is well known as well as the density of electrons as a function of time (Peebles 1993), giving $\nu \sim 4{\times}10^{26}$ $m^2$ $s^{-1}$ (Gibson 1996). The initial fragmentation mass scale of the plasma is easily calculated to be that of superclusters.

Clusters and galaxies form as the plasma cools. Voids grow at sonic speeds $c/3^{1/2}$ creating weak turbulence at void boundaries (Gibson 2010). Because Reynolds numbers of the turbulence is small, the Schwarz turbulence scale $L_{ST} = \varepsilon^{1/2}/(\rho G)^{3/4}$ is not much larger than $L_{SV}$, where $\varepsilon$ is the viscous dissipation rate of the turbulence. Thus the Kolmogorov scale $L_K = (\nu^3/\varepsilon)^{1/4}$ of the weak plasma turbulence fixes the size and mass of the protogalaxies emerging from the plasma epoch with the Nomura weak turbulence morphology, where all quantities are known. The kinematic viscosity of the gas is much smaller than that of the plasma at transition, so the viscous Schwarz scale is decreased and the mass scale is that of the Earth. Fragmentation also occurs at the Jeans scale because heat is transferred at light speeds but pressure is transferred at sound speeds. This mismatch makes it impossible to maintain constant temperature as primordial gas fog particles PFPs fragment.

## 4. Conclusions

Gravitational structure formation after the turbulent big bang occurs in the plasma epoch by a viscous fragmentation process beginning at 30,000 years, or $10^{12}$ seconds. Protosuperclustervoids expand at near light speeds starting at this time to form the $10^{25}$ meter completely empty regions observed at present, contradicting the standard $\Lambda$CDMHC cosmology that suggests voids are the last features to form as the universe evolves rather than the first.

Weak turbulence results from the plasma void expansions that determines the morphology and scales of the protogalaxies, as shown in Figs. 4-6. The kinematic viscosity of the primordial gas $\gamma \sim 10^{13}$ m$^2$ s$^{-1}$ at the transition temperature $\sim$ 3000 K. Protogalaxies promptly fragment at Jeans mass to form PGCs and Earth mass to form PFPs, as shown in Fig. 5 (top).

All stars form by PFP mergers within PGC clumps. Most of the PFPs remain as frozen hydrogen gas planets. Clumps of PGC clumps comprise the dark matter of galaxies. The non-baryonic dark matter diffuses to form galaxy cluster halos, as shown by Fig. 7. We see that rather than $\sim$ 10 planets per star there are 30,000,000. It is easy to understand Jupiter mass planets observed orbiting stars at distances matching the orbit of planet Mercury using hydrogravitational dynamics HGD cosmology. It is easy to understand overfeeding stars to form supernovae by continued planet accretion. Gravitational collection of the chemicals produced by the hydrogen gas planets reduces iron and nickel to metals, produces rock layers, oceans and the complex chemistry of life.

It seems clear that the $\Lambda$CDMHC cosmological model should be abandoned.